# Beyond opening up the black box: Investigating the role of algorithmic systems in Wikipedian organizational culture

R Stuart Geiger

## Abstract
Scholars and practitioners across domains are increasingly concerned with algorithmic transparency and opacity, interrogating the values and assumptions embedded in automated, black-boxed systems, particularly in user-generated content platforms. I report from an ethnography of infrastructure in Wikipedia to discuss an often understudied aspect of this topic: the local, contextual, learned expertise involved in participating in a highly automated social–technical environment. Today, the organizational culture of Wikipedia is deeply intertwined with various data-driven algorithmic systems, which Wikipedians rely on to help manage and govern the "anyone can edit" encyclopedia at a massive scale. These bots, scripts, tools, plugins, and dashboards make Wikipedia more efficient for those who know how to work with them, but like all organizational culture, newcomers must learn them if they want to fully participate. I illustrate how cultural and organizational expertise is enacted around algorithmic agents by discussing two autoethnographic vignettes, which relate my personal experience as a veteran in Wikipedia. I present thick descriptions of how governance and gatekeeping practices are articulated through and in alignment with these automated infrastructures. Over the past 15 years, Wikipedian veterans and administrators have made specific decisions to support administrative and editorial workflows with automation in particular ways and not others. I use these cases of Wikipedia's bot-supported bureaucracy to discuss several issues in the fields of critical algorithms studies; critical data studies; and fairness, accountability, and transparency in machine learning—most principally arguing that scholarship and practice must go beyond trying to "open up the black box" of such systems and also examine sociocultural processes like newcomer socialization.

This article is a part of special theme on Algorithms in Culture. To see a full list of all articles in this special theme, please click here: http://journals.sagepub.com/page/bds/collections/algorithms-in-culture.

## Introductory vignette: The professor's simple question

As a long-time volunteer Wikipedian editor (I first began editing in 2004) and as someone who has studied the social and organizational dynamics of the "anyone can edit" encyclopedia project for many years, I am frequently asked for advice from people want themselves or their organizations to be represented in particular ways on Wikipedia. This introductory composite vignette[1] is about one such case, with a professor who had some concerns with how they were represented on Wikipedia. They told me the article others

Berkeley Institute for Data Science, USA

**Corresponding author:**
R Stuart Geiger, Berkeley Institute for Data Science, Berkeley, CA 94720, USA.
Email: stuart@stuartgeiger.com

wrote about them was factually accurate, but there were some things the professor thought should be rephrased. They asked me how someone in their position ought to go about correcting their own Wikipedia article, in line with the community's rules and norms. This professor knew quite a bit about Wikipedia—enough to look through the revision history and learn who added the awkward phrases and when, and enough to know that Wikipedia had rules and norms that might make it unwise to edit the article themself. Yet like most people who ask me for advice on editing Wikipedia, they were not certain about the community's increasingly complex labyrinth of policies and processes to know for sure how to proceed. They wanted to know more about what would be involved. In line with Wikipedia's policies and values, when I receive these kinds of requests, I do not edit on someone else's behalf, but instead help teach and empower them to participate in Wikipedia themselves. When I received the professor's e-mail, I got excited and sent a long reply on how exactly they should proceed if they did want to make a change in line with Wikipedia's rules and norms.

*The top of the edit request list as of 19 May 2015, with 87 requests pending*

In short, they could edit it themselves, but best practice would be to get at least one Wikipedian—an active volunteer editor who had experience in the community—to approve the requested changes before making them to the article. However, the process to properly get such approval would take a few steps. First, they needed to register an account that identified them, as Wikipedians prefer that people with conflicts of interest declare them outright. Then, they needed to go to the designated "talk" page for the article, edit the talk page, and add a new message. They should identify the issue and suggest what he thought was an appropriate change. Finally, I recommended adding a special trace to make sure a veteran Wikipedian editor saw that request, since it might take weeks before someone stumbled onto the article's not-so-active talk page. So, at the top of the message requesting particular changes to the article, they needed to add the text `{{request edit}}`. This special text (called a template) was not intended for humans, but instead written to get the attention of an automated software agent named AnomieBot, which was continually scanning for every new talk page message in near real time for this text. In a matter of minutes, when the bot's script next ran on a server cluster in Ashburn, Virginia, it would summarize the professor's edit request and put it into a centralized queue in a special administrative space in Wikipedia, alongside all the other pending requests of this same type. The bot would provide some information about the request, such as the date of the request and if the page is locked down from public editing, which it displays more prominently in yellow.

Sometime later, a Wikipedian who was looking for something to do would check that list, navigate to the talk page for the professor's article, and give their thoughts about whether the edit request was a good idea or a bad idea. This person would most likely be someone who spent a lot of their time on the English-language Wikipedia responding to conflict of interest edit requests, rather than any number of other tasks they could possibly do. (There is substantial specialization and division of labor among Wikipedians, in part supported through these kinds of centralized venues.) If the Wikipedian agreed with the professor (which would be likely, I thought) they might edit the article themselves to fix it. Or they might tell the professor to implement the changes on their own. However, if the Wikipedian didn't agree, they would explain why, and the two might have a longer discussion on the article's talk page. With any decision made, the Wikipedian should leave a slightly different standardized template based on what their decision was, so that AnomieBot would remove the professor's edit request from the queue and place it into an archive.

I was quite enthusiastic about explaining this highly structured and automated administrative process, hoping that they would find it as fascinating as I did. Perhaps they would even start to participate in the Wikipedia community on other articles in their area of expertise. However, their reaction couldn't have been more different to mine. The professor—as someone with substantial experience authoring many different kinds of reference works, who simply wanted to have some say about how the Wikipedia article about them was worded—was immediately disinterested by the amount of effort that would be involved. My first reaction to their reaction was to pontificate on the principles and practicalities underlying this system—the reasons why this kind of template–bot–noticeboard–veteran workflow was implemented, rather than any number of alternatives that might seem more practical on the surface. But I realized that the more interesting question is why I had so internalized this socio-technical assemblage and the values it enacts. I have spent over 10 years participating in this particular social–technical system, where these kinds of bot-supported processes and workflows are a routine, taken-for-granted part of what it means to work in Wikipedia. They are deeply imbued with particular values, principles, norms, and ideals, and learning them is not just about technical competency, but also socialization into a complex organizational culture—one that can be quite different than other professionalized cultures of knowledge production.

## Article overview

### Bots in Wikipedia: An understudied data consumer of the Wikipedia corpus

Wikipedia, as one of the world's largest and most visited sites of knowledge production and dissemination, is frequently used as a corpus of "big data" for text mining on a variety of topics.[2] In this article, I focus on an understudied data consumer of the Wikipedia corpus: the volunteers who author, edit, and maintain the encyclopedia project, who have increasingly developed and relied on data-driven algorithmic systems for helping manage their work at enormous scale. Like in any large-scale organization, Wikipedians spend a substantial amount of time on the "meta work" involved in reviewing, managing, coordinating, and policing the hundreds of thousands of people who help write and edit Wikipedia (Kriplean et al., 2008; Wattenberg et al., 2007). Each language version has dozens of distinct processes for making decisions, distributing tasks, and resolving conflicts at a variety of scales (Forte and Bruckman, 2008). And just like many companies, governments, and other organizations that have to make decisions at scale, Wikipedians have been increasingly turning to automation, expert systems, artificial intelligence, and other algorithmic systems to help with this meta work.

Today, thousands of fully and semi-automated software agents like AnomieBot operate in and across the various language versions of Wikipedia, programmed to carry out particular tasks that are needed to help keep Wikipedia running smoothly. AnomieBot's edit request code is relatively simple, as is the code powering many other Wikipedia bots, which is built on top of an infrastructure maintained by the Wikimedia Foundation that lets developers easily parse through every edit made to Wikipedia in near real time. Together, these thousands of agents have profound impacts on how Wikipedians accomplish the work of writing and editing an encyclopedia. In the English-language Wikipedia, 22 of the 25 most active editors (by number of edits) are bot accounts, and July 2017, they made about 20% of all edits to encyclopedia articles.[3] Scholars in this area have argued that Wikipedia's unexpected success in the face of few top-down management structures is not due to a "wisdom of crowds," but rather a wisdom of bots, algorithmic agents, and smart interfaces that help structure the work of coordination behind the scenes, in the absence of formal, top-down management (Niederer and Van Dijck 2010; Geiger 2011a; Halfaker and Riedl, 2012).

While some of these bots have been studied for their role and impact in the Wikipedian community, we must also look to how the lived experiences of being and becoming a Wikipedian has changed in an increasingly algorithmized organizational culture. Bryant et al. (2005) discussed how part of becoming a Wikipedian was learning how to use the wiki platform, but since their 2005 study, Wikipedians have adopted a substantial suite of bots, tools, scripts, extensions, and dashboards. What does it mean to contribute to "the free encyclopedia that anyone can edit" when that participation requires not only learning Wikipedia-specific jargon, norms, style guides, and rules, but also learning how to interact with all the bots and power tools that veteran Wikipedians rely on to manage, track, triage, and coordinate the different kinds of administrative and meta work that goes on behind the scenes? Ford and Geiger (2012), for example, discuss various changing organizational literacies in Wikipedia, including cases where newcomers were unsure about whether they were interacting with a human or a bot. Extending this, I argue that the bots, automated tools, and encoded routines that undergird participation in Wikipedia are as much a part of Wikipedia's particular organizational culture as more traditionally social or cultural elements like jargon, norms, ideologies, epistemologies, argumentative styles, subcultural configurations which have been extensively studied by sociocultural researchers and are continually reflected on by Wikipedians themselves.

### Algorithms as culture/algorithms in action

While my research focused on the English-language Wikipedia, my argument about the role of algorithmic systems in groups, organizations, and cultures has broad implications. This research topic is related to a broader trend across the public and private sector, as data-driven algorithmic systems are increasingly deployed to organize, order, govern, and gatekeep social networking and social computing platforms such as Google, YouTube, Facebook, Twitter, Uber, and AirBnB. Such systems are also playing significant roles in traditional social, political, and economic institutions, including areas such as predictive policing, algorithmic sentencing in criminal trials, credit scoring, scores and benchmarks for hiring and firing workers, and systems that determine eligibility for and allocation of social services. Scholars from a variety of fields are increasingly studying the roles that such systems, practices, and approaches have in our world (boyd and Crawford, 2012; Grosser, 2014; Nakamura, 2013; O'Neil, 2006; Pasquale, 2015; Thrift and French, 2008; Tufekci, 2014). The specific computational and statistical techniques deployed in these systems differ dramatically, but there are many shared issues and implications about systems encode procedures in a social, cultural, and/or organizational context.

There is a growing body of scholarship aligning under the title of "critical algorithms studies" (Gillespie and Seaver, 2016; Seaver, 2013) to investigate these issues. Furthermore, concerns about the role of software in the governance of online communities has been a core topic of the software studies field for decades, best encapsulated in Lawrence Lessig's (1999) famous declaration that "code is law," speaking to the governmental role that programmers have in designing and developing systems. Scholars from history, philosophy, and the social study of science and technology have long investigated the ways in which technologies of knowledge production have cultural, political, and economic implications. A classic text in this lineage is Max Weber's analysis of the crucial role double-entry bookkeeping played in the establishment of bureaucracies and modern capitalism (Weber, 1978 [1922]), with contemporary scholars examining many issues around the deployment of automated systems in workplaces (e.g., Orlikowski and Scott, 2008; Ribes et al., 2013; Yates, 1989).

Contemporary scholars investigating these systems have extensively focused on their proprietary, "black-boxed" source code, as they are making decisions that have wide-reaching impact with little public accountability. I agree that proprietary source code is an important barrier to openness and public accountability, but it is far from the only one. In the case of Wikipedia, which has a strong open source ethos, we get a glimpse into a potential future world in which platforms' key algorithmic systems are open sourced by default. In this context, Wikipedia demonstrates how the issues in and around algorithmic systems are as much social as they are technical, going far beyond the opacities that arise around proprietary source code. My argument extends Burrell's (2016) discussion of three different forms of opacity in machine learning: intentional secrecy (proprietary source code), technical literacy (such as learning to read code), and opacities inherent in machine learning (such as issues of interpretability). To these forms, I add another: the opacities in learning a particular institutional or organizational culture that is supported by algorithmic systems.

I define "algorithmic" as involving encoded procedures, which are typically—but not exclusively—computationally implemented. Following Seaver (2013), I focus not on "algorithms" in the abstract but on "algorithmic systems"—how these encoded procedures are deployed in a complex social–technical context. The code powering AnomieBot was written to support a particular organizational process undergirded by social norms and ideological principles, and in doing so, the bot further supports the process, norms, and principles. If AnomieBot was using advanced neural network classifiers to identify edit requests to be aggregated (rather than simple pattern matching), there would be an additional layer of algorithmic complexity at work, but my discussion would still be how an organizational task is abstracted such that it can be "clerked" by an automated software agent.

I also take from Seaver's (2017) call to study "algorithms as culture" rather than "algorithms in culture"—focusing on the co-construction of culture through a variety of means, including but not limited to algorithmic systems. My goal in this article is to relate a more empirically situated and constructivist approach to studying algorithmic systems "in action" or "in the wild," much as scholars in the field of science and technology studies have investigated science and technology (e.g., Hutchins, 1996; Latour, 1987). This approach extends the "Critical Data Studies" approach, which Iliadis and Russo (2016) describe as "captur[ing] the multitude of ways that already-composed data structures inflect and interact with society, its organization and functioning, and the resulting impact on individuals' daily lives." Scholars in this area have similarly focused on topics like "how digital data become meaningful in mundane contexts of everyday life" (Pink et al., 2017), how data scientists think about what makes a good or bad process for data analysis (Lowrie, 2017), and the situated perspectives and "data ideologies" that publics have about open government data initiatives (Schrock and Shaffer, 2017). Just as our understanding of "big data" and "data science" benefits from understanding the lived experiences of data in people's local, situated contexts, so does our understanding of the algorithmic systems that operate on large-scale sources of data.

## Methodological approaches to studying algorithmic systems "in action"

### Method and fieldwork

This paper is based on my multiyear ethnographic engagement in the culture, organization, governance, and infrastructure of Wikipedia, which began with a deceptively simple question: when I read an article on a topic like evolution, Harry Potter, the U.S. Civil War, or Manchester United, why does it represent the topic in the way it does, rather than any of the many alternative ways that people around the world could imagine? In short, how is the "anyone can edit" encyclopedia moderated? I have participated in the Wikipedian community as a volunteer in various capacities for 12 years, starting in 2004. I started more formally studying Wikipedia as an ethnographer in 2008, conducting participant-observation, interviews, archival analysis, and computationally supported social science research. My fieldwork involved participant-observation and interviews, generally in Wikipedia and other mediated platforms Wikipedians use (mailing lists, Internet Relay Chat rooms), but also in meetups in colocated spaces. I engaged in thousands of hours of editorial work in Wikipedia articles, discussions, and policies, which (as I detail in this paper) is heavily algorithmically assisted. I also specifically worked in Wikipedian bot development, including proposing, developing, and operating a bot of my own, as well as helping other bot and tool developers redesign their bots and tools. I attended dozens of Wikipedia-specific colocated events, including meetups, edit-a-thons, conferences, and hackathons, and interned at the Wikimedia Foundation for three months. I also conducted extensive archival analyses of the history of Wikipedia, focusing heavily on the development of governance structures and the project's software infrastructure.

## Telling stories about contextual factors

Like all algorithmic systems, the ones I studied in Wikipedia were designed, developed, and deployed by people. Like other scholars, I find it important to give rich, thick descriptions about how and why these systems were created and maintained as they were (Arnoldi, 2016; Geiger, 2014; Hallinan and Striphas, 2016; Kitchin and Dodge, 2008; Mackenzie, 2013; Seaver, 2013). As Gillespie (2014) argues: "A sociological analysis must not conceive of algorithms as abstract, technical achievements, but must unpack the warm human and institutional choices that lie behind these cold mechanisms." Like any social institution, Wikipedia is certainly an ongoing accomplishment that takes work to maintain, and people work to maintain it in certain ways and not others. Algorithmic systems come on the scene in Wikipedia not only as ways for people to increase the efficiency of certain tasks, but also as a way for people to advance particular notions about Wikipedia's organizational structure and culture, as scholars of IT in organizations have argued in different kinds of workplaces (Orlikowski and Scott, 2008; Star and Strauss, 1999; Zammuto et al., 2007).

Examples of previous kinds of sociological studies of the broader implications of algorithmic systems include Introna's (2016) discussion of the plagiarism detection software Turn It In using Foucault's concept of governmentality. Similarly, Crawford and Gillespie (2014) analyzed interfaces in major social networking and social media sites developed for users to flag or report inappropriate content. They argue that such interfaces are articulations of a "vocabulary of complaint" that structures a highly automated human–computational system used for moderation work, the inner workings of which are opaque to all but a few who work for Facebook, YouTube, Twitter, etc. People who use a platform align their expectations and actions to this vocabulary of complaint, which also delineates what is and is not knowable by those who work for the platform as moderators.

Yet Seaver (2013) notes how most scholarship in the emerging field of "critical algorithms studies" continually struggles with the inability to "see inside the black box" of algorithmic systems, which makes it difficult to see such systems "in the making." To see a system "in the making" is the classic denaturalization move continually made by ethnographers and historians of science and technology who investigate the messy work and competing perspectives that are often obscured when a product goes to market or a paper goes to a scientific journal. In his reflections on ethnographically studying the people who develop music recommendation systems, Seaver (2013: 9–10) notes how his own thinking changed as a result of his fieldwork, in which he was able to enter and observe the sociotechnical systems in which algorithms were being designed, developed, and deployed:

> It is not the algorithm, narrowly defined, that has sociocultural effects, but algorithmic systems — intricate, dynamic arrangements of people and code... When we realize that we are not talking about algorithms in the technical sense, but rather algorithmic systems of which code strictu sensu is only a part, their defining features reverse: instead of formality, rigidity, and consistency, we find flux, revisability, and negotiation.

In this context, I ask: for whom are algorithmic systems (and the organizations that rely on them) formal, rigid, and consistent, and for whom are they in flux, revisable, and negotiable?

## A methodological analogy: First- versus second-level digital divides

A core principle of my ethnographic research holds that, if we are concerned with how a heavily automated organization of knowledge production like Wikipedia works, we must pay attention to both the code and culture that makes up a broad sociotechnical system. The two vignettes I present illustrate these contextual factors in the case of Wikipedia, showing that Wikipedia's bot-assisted bureaucracies are quite different for newcomers versus veterans. In one sense, this argument is a kind of algorithmic version of foundational lessons learned in the sociological study of digital divides. As Eszter Hargattai (2002) argues in her analysis of "second-level digital divides" on the issue of inequality, the first level of digital divide discourse focused on who had access to the Internet. As more people got access to computers and the Internet, there was a different source of inequality, driven by who had the knowledge, skills, and sense of empowerment to use the Internet in ways that further engaged, empowered, and benefitted their lives.

I argue for a similar reframing around the issue about who influences the content of Wikipedia articles in a time when algorithmic systems play substantial roles in the project. We must look at more than the fact that participation in Wikipedia is open to the public; that the infrastructure supporting it is open sourced; and that the community's policies, procedures, and norms are documented in thousands and thousands of pages of text. We must also look at what kind of skills, knowledge, and investment is required to fully and successfully participate, particularly examining the roles that algorithmic systems play in raising the complexity of participation. My work is therefore aligned with studies that have examined how Facebook users understand the algorithmic filtering of the news feed and, in some cases, learn enough to change their practices and expectations accordingly (Bucher, 2016; Eslami et al., 2015), although in a quite different context due to Wikipedia's particular stance toward open source software and open participation. Yet the difference is a productive one: we can imagine a hypothetical world in which all the work done by Facebook's employees on the news feed are done publicly in an open source code repository (and an open issue-tracking/bug-reporting platform, as Wikipedia also has). In this world, what kinds of people would and would not have the time, expertise, inclination, and sense of empowerment to dive into such code and documentation?

## Vignette two: The speedy deletion process

In this section, I relate a second ethnographic vignette about an experience I had in Wikipedia, working on an article for a topic I had chosen to write about. This vignette illustrates how a quite different bureaucratic process—around the deletion of substandard encyclopedia articles—is suffused with different kinds of algorithmic agents than the previously discussed edit request process.

Today, participating in Wikipedia involves working alongside bots, which are designed to read certain kinds of digital traces that align with particular bureaucratic processes. However, like all bureaucracies, these bot-assisted processes can operate with quite different priorities and assumptions, even though they have some similarities and common elements. The ability to properly make a conflict of interest edit request in Wikipedia doesn't necessarily translate to ability to challenge the deletion of an article, any more than learning how to apply for a driver's license gives someone the capacity to defend themselves in a criminal trial. Navigating these processes requires substantial social and technical literacy, which I have obtained as a long-term Wikipedia contributor. For veterans, interacting with bots and semiautomated tools is often a taken-for-granted affair—some bot developers have told me that their bots work so seamlessly they are concerned other Wikipedians may not even know there is a bot automating various aspects of their workflow.

Like the edit request process, the speedy deletion process is also driven by adding templates to wiki pages, which programmatically leave a prewritten block of text on an article and performatively serve as actions asserted by Wikipedians in this specific bureaucratic procedure. To nominate an article for speedy deletion is to add a template like `{{db-web}}` to the top of an article, and vice versa, making it a kind of performative utterance (Austin, 1976). Similarly, to properly contest the nomination of an article for speedy deletion is to add, edit, or remove various other templates in particular ways (which differ based on the role that each editor has in this process—nominator, article creator, administrator, or a third party). There are semiautomated browser extensions and other "bespoke code" (Geiger, 2014) that veterans enable and install, which scaffold and structure this work ought to take place. As this vignette shows, the regularity of this process is articulated in written textual policies that are grounded in abstract ideals, but implemented in various software programs and enacted by the people who participate in this algorithmized workflow.

### "Save saveMLAK!"

On the morning of 9 August 2013, I was sitting in the auditorium of Hong Kong Polytechnic University, where the opening ceremonies of Wikimania 2013 were being held. Wikimania is an annual hybrid convention/conference for those active in Wikimedia Foundation projects (including but not limited to Wikipedia). The keynote speaker was Makoto Okamoto, the founder of a wiki called saveMLAK, dedicated to coordinate responses and efforts to the 2011 Tohoku earthquake and tsunami. With my laptop out, I looked for an article on saveMLAK on the English-language Wikipedia, finding none. So I did what Wikipedians do and created one, starting off with a barebones "stub" and hoping that others would expand it during the keynote.

At 9:43 am local time, I created the saveMLAK article, which was expanded by another Wikipedian (also in attendance) three minutes later. I started adding an "infobox," but when I went to add this at 9:47 am, I found someone had used a semiautomated tool named Twinkle to put a giant red notice on the article. This prewritten template, invoked by adding {{db-web}} to the text of any Wikipedia article, is part of the "criteria for speedy deletion" (CSD) process. Speedy deletion is one of three different processes Wikipedians have for deleting articles from Wikipedia. Twinkle is a browser extension that approved Wikipedians can use to easily insert these prewritten templates, authoring all the discourse needed to participate in this standardized bureaucratic process. Anyone can manually add the prewritten nomination text, but Twinkle scaffolds, routinizes, and streamlines this process. Wikipedians who use algorithmically assisted tools and extensions like Twinkle are best understood as cyborgs (Halfaker and Riedl, 2012), with their experiences and affordances extended in particular ways and not others.

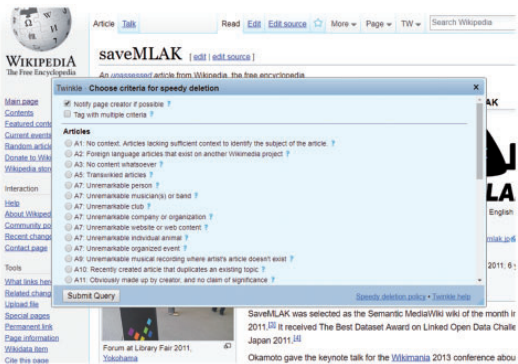

How a Twinkle user would request speedy deletion of a page. Clicking the TW tab opens up a drop-down menu, clicking CSD pops up a dialog box, where one of the rationales is selected.

According to long-established policy and process (which is enacted in these templates, bots, and tools), any editor can "CSD" or "speedy" an article they believe fits one of several dozen criteria by tagging it with certain templates. These templates generate the text Wikipedians used to manually write when arguing for deleting an article according to the project's notability policies, and the templates make such nominations visible to a large set of human and algorithmic users who know how to follow this trace. The {{db-web}} template (rendered above) was left on the SaveMLAK article, contains text arguing that the article about a website fails the A7 criteria in the CSD process, which demands that articles "credibly indicate the importance or significance of the subject." (A majority of speedy deleted articles are tagged with templates containing A7 rationales (Geiger and Ford, 2011).) As part of the CSD process, those who tag articles for speedy deletion are supposed to notify the original author of this so they can properly respond, which is a process that has also been automated by tools like Twinkle that structure the workflows of Wikipedians who engage in this kind of quality control work. Accordingly, a few seconds after I saw the tag on the SaveMLAK article, I received a new message on my talk page, prewritten but signed by the Wikipedian who tagged my article for speedy deletion.

Once an article is tagged in this way, it will then be automatically aggregated by software agents to a few different centralized spaces where administrators can review articles that have been recently CSDed. Administrators have the technical privilege in the software to unilaterally delete (or undelete) any page, but the CSD process follows a "four eyes" principle (like many in Wikipedia). Administrators are only authorized to delete articles if someone else has first independently evaluated it and deemed it worthy of speedy deletion, indicated by tagging it with a CSD template. However, if two Wikipedians believe that the article should not be speedy deleted, then that is considered sufficient cause to take it out of the CSD process, possibly putting it instead in the more rigorous Articles for Deletion process. This ideal is implemented in practice in that anyone except for the article's original creator is allowed to remove the CSD template tag without discussion or justification. Administrators are not supposed to delete pages that have had the tag removed by someone who is not the article's original creator. The article's original creator is

The prewritten speedy deletion/CSD notice on the SaveMLAK article, 9:47 am

*The prewritten speedy deletion notice I received for the SaveMLAK article*

presumed to always agree that the page should not be deleted, so they are not allowed to remove the tag; doing so can result in admonishment and a temporary block if repeated.

When I saw the CSD A7 notice appear on the page, my heart sunk. I didn't even have to read it, as I knew exactly what it said. I'd seen the same notice thousands of times before, and I'd even helped rewrite one of its related templates to make it more user friendly. I also knew exactly what I had to do, and that I might only have seconds to do it. As the article's original creator, I couldn't legitimately remove the deletion nomination tag, but I could add another template-based tag—{{hangon}}—that would signal two things to any administrator going through the speedy deletion process. The first signal was more explicit, telling them that I was actively working on expanding the page. After adding this template, a prewritten message saying as much would appear at the top of the article and would be visible on their screens if they were using most of the popular in-browser and stand-alone bespoke tools that Wikipedians have developed to automate various parts of this process. Yet I had a second, subtler motivation, hoping that in properly demonstrating correct usage of such a template within the established workflow of this process, I would be made legible as a Wikipedian who knew the CSD process and should be given some more leeway—unlike most of the people who were creating articles that they were deleting.

As I added the {{hangon}} tag in the proper place and clicked the "submit" button at 9:48 am, an error message appeared in my browser: the article I was editing no longer existed, as it had been deleted. "Of course," I thought. Below this error message, I saw the "Start the SaveMLAK article" link that would let me recreate the article if I so desired, but I knew that would be the last thing I should do at this moment. I needed to get an administrator to undelete the article, or else the recreated article would be CSDed again and possibly "salted"—when the deleted page is protected from editing so no non-administrators can create a new version. Normally, what a non-administrator like me would do is go through the Deletion Review (DRV) process, where I'd write up my case, submit it, have it enter a queue, wait for an administrator to process it, have some back and forth with them, and so on. But I was in a thousand-person auditorium filled with Wikipedia's upper echelon, all listening to a captivating presentation about this article that just got deleted. Hundreds of people in the room had the technical privilege to undelete the article using their administrative accounts, and any one of them would be authorized to unilaterally do so without needing to even give much of a justification, given the procedures specified in the CSD policy. Had the article been deleted through the more rigorous Articles for Deletion process, an administrator could only undelete it after it went through Deletion Review, but any admin can reverse a CSD. So I posted about the deletion to Twitter with the #Wikimania hashtag attendees had been using. One of the many administrators in the audience saw the tweet and promptly undeleted the article in accordance with the CSD process at 9:51 am, three minutes after it was deleted. Another Wikimania attendee notified the admin who deleted the article about the undeletion on their user talk page (not required, but done as a matter of courtesy), and a polite discussion took place there. The deleting admin stated they were authorized under policy and process to have deleted the saveMLAK article when they did, but agreed that it might be better to wait more than a few minutes before deleting new articles. I replied, telling the deleting admin that they could nominate it at AfD if they felt it was still underdeveloped in a few hours.

## Discussion

### Generalizing the ethnographic vignettes to the rest of Wikipedia

My aim with the two vignettes in this paper is to illustrate how participation in Wikipedia has increasingly involved interacting with not just a single automated system, but complex, overlapping networks of automated systems and bureaucratic procedures. I have chosen these two vignettes for multiple reasons: they both take place hundreds of times a month, they are representative of my experience participating in Wikipedia (they are not abnormal outliers), and these stories echo ones I have heard from newcomers and veterans alike in interviews. I chose the edit request and speedy deletion processes out of the dozens of others because they illustrate the diversity of processes. Speedy deletion is fast-paced, high-volume, high-stakes, demands administrator intervention for each case, and is based on a policy with a highly complex ruleset. The edit request process is slower, has less volume, is generally lower stakes, does not require administrators to resolve in most cases, and is based on one of Wikipedia's less complex policies (which is still complex). However, both speedy deletion and edit requests rest on a similar assemblage: ideals and values are implemented in policies and rules, which define an organizational process, which is encoded in software that is built to help support Wikipedians.

While I have only given thick descriptions of two algorithmically assisted bureaucratic processes in Wikipedia, the community's ecosystem of fully automated bots and semiautomated power tools extends far beyond those two cases. Wikipedia is a rich site for studying how groups, communities, and organizations develop, design, deploy, and debate algorithmic systems. There are thousands of bots in Wikipedia, and tasks that they are delegated extend to every level of the encyclopedia and the community who writes it. Bots have existed almost from the beginning of Wikipedia's 15-year history, starting with Ram-Bot in 2002, which almost doubled the size of Wikipedia by creating an article about every city and town in the U.S. from public domain census data (Livingstone, 2016). A suite of different bots, maintained by different contributors, automatically remove edits that they determine to be spam, vandalism, plagiarism, or gibberish (Geiger and Halfaker, 2013). Furthermore, hundreds of bots like AnomieBot work to support particular organizational processes, "clerking" for a particular task, like fact-checking statements, resolving disputes between Wikipedians, or deciding what content should be featured on the main page (Gilbert and Zachry, 2015). Because of this, much of the relatively high quality and internal consistency of Wikipedia should be attributed as much to "wisdom of bots" as is attributed to the frequently cited "wisdom of crowds." As the project has scaled from a small group of tight-knit contributors, Wikipedians developed a wide variety of bots to do much of the routine, mundane work that is needed to support Wikipedia as both an encyclopedia and a community.

Such systems do not eliminate the need for human labor, but instead transform the kind of work that takes place, as well as what it means to be a Wikipedian and participate in this community. Stories like those of the professor's request or of the deletion of my article are not outliers; they take place every day in and around Wikipedia. In my own experience trying to help many different kinds of people learn how to edit Wikipedia, an overwhelming majority eventually give up because they are frustrated with both the social and technical learning curve required to participate as Wikipedians expect. Furthermore, my ethnographic findings resonate with several quantitative analyses of participation in the English-language Wikipedia. One analysis showed that in 2011, 70% of articles that were nominated for speedy deletion were tagged within 10 minutes of being created, and 25% of articles nominated for speedy deletion were deleted within 10 minutes of being created (Geiger, 2011b). Quantitative studies have also investigated the socialization and attrition rates of new contributors in relation to these changing sociotechnical practices, finding that out of the newcomers who made high-quality contributions, 25–35% continued to edit two months after their first contributions in 2003–2006, but after the introduction of these new tools and processes around 2007, that figure dropped to around 10% in 2007–2009 and 5% in 2010 (Halfaker et al., 2013). The same study showed that by 2011, around 40% of newcomers making high-quality contributions had at least some of their edits reverted by fully automated bots or humans using semiautomated power tools, which had significantly contributed to the size of the editing community dropping from a peak of over 55,000 highly active editors per month in 2007 to a trough of under 35,000 highly active editors per month in 2012. A more situated view like the one in this article helps give context to these quantitative findings.

## What do we learn from taking a more situated view?

*Technical decisions are sociocultural decisions and vice versa.* For non-Wikipedians who hear of stories like the deletion of the saveMLAK article, the speed of interactions is often striking. An impact/effects approach to studying automation might discuss how bots and tools have made this process more efficient, such as the previously cited literature discussing the impact of automation on participation. My more situated approach to this issue explores how automation is a way in which speed is further woven into the fabric of Wikipedia's culture, but not universally. Speedy deletion is speedy for a particular reason based on the values of Wikipedians in the community who first created and formalized it in 2004–2006, when Wikipedia was a frequently ridiculed novelty in the public imagination. The speedy deletion process was intentionally developed to be a more efficient alternative to removing low-quality content than the slower Articles for Deletion process, which involved consensus building over a seven-day period before a decision would be reached. The edit request process is similarly techno-bureaucratic, but with a different cultural history, arising out of a different set of concerns around conflict of interest and beliefs about neutrality, and as such, it has a different temporal rhythm.

Decisions about how Wikipedia ought to operate are made in and through the design and deployment of fully automated bots and semiautomated tools. The resulting sociotechnical structures are specifically built upon a set of shared beliefs and understandings, such as in the procedure of speedy deletion. The formal CSD are rules, written in English text to help Wikipedians collectively make quick decisions about whether an article ought to be deleted. The speedy deletion process is formalized in a policy that is also written in English text, to help Wikipedians collectively interpret and apply these criteria in a standardized way. Yet the speedy deletion process is also formalized in the bots and tools that have been built to help automate, streamline, and standardize participation in this particular process. The abstract ideals of the process—such as the "four eyes" principle, holding that no article ought to be deleted unless one administrator and one other Wikipedian editor both believe it fits the speedy deletion criteria—are collectively enacted through the software programs that are constantly querying recent changes to Wikipedia to support this workflow.

*How people differently relate to the same sociotechnical system?* Thick descriptions of average, everyday routines in algorithmized contexts help us keep both the social and technical aspects of algorithms in view. As Seaver (2013) notes with his critiques of various "critical algorithms studies" literature, it is easy to slip into a mode of analysis where social factors are contextualized, while infrastructure remain static and determining. Such an approach "keeps algorithms themselves untouched, objective stones tossed about in a roily social stream" (10). A critical social science study of algorithms ought to argue that both people's understandings of algorithmic systems and the algorithmic systems themselves are constructed, negotiated, contextualized, and differently interpreted and enacted. Studying socialization, literacy, and newcomer–veteran interactions are particularly powerful ways of keeping this dual constructivist view in mind. It is a false question to ask whether algorithmic systems in general are more or less negotiable than social systems in general. Rather, we ought to be asking: for whom are these sociotechnical systems negotiable and what do people have to do in order to exercise personal agency?

For example, understanding the edit request process involves learning several layers of automated systems and information infrastructures. It involves learning that there are templates, invoked by bracketed text like {{edit request}}, that will render as different kinds of display elements when added to the text of wiki pages. It also involves learning that if you leave certain templates on a wiki page, a bot will be summoned to do some task based on how you called the template—like how {{edit request}} summons AnomieBot to put an edit request in the centralized queue page. It also involves learning that when you leave certain templates on a page that are indexed in various ways, other people on Wikipedia will be able to find it based on what they are looking for. And it also involves learning that these templates link up with formalized policies and procedures defining how Wikipedians are supposed to act and make decisions.

In the case of speedy deletion, the decision to add {{db-a7}}, {{citation needed}}, or {{hoax}} to an article sets off a chain of events that will be interpreted within Wikipedia's particular sociotechnical context. Veterans know precisely what will happen if they leave one of these templates in the right place—or they know how to find out what will happen, or they know how to deal with the consequences if they make a mistake, and so on. Veterans also know how to tell if others have properly participated in a given process, and many veteran Wikipedians I have interviewed have told me about the ways in which they determine if someone has performed algorithmic-bureaucratic competency, similar to how Ford and Geiger (2012) discuss the "trace literacies" involved in participating in Wikipedia. I have found that my own participation in Wikipedia has shifted accordingly, where I often find myself performing these trace literacies in the hopes of making myself legible as a Wikipedian.

*Transparency as open software code versus algorithmic literacy.* Wikipedia's computational infrastructure is also designed and governed in a relatively open manner by the project's volunteer community of editors (Forte and Bruckman , 2008; Gilbert and Zachry, 2015; Kennedy, 2010), unlike most of the automated systems that are increasingly prevalent in digitally mediated environments. There is a formal bot policy and bot approval process (Geiger, 2011a), which functions similarly to Wikipedia's content policies for encyclopedia articles. Wikipedians discuss and debate about what kinds of bots should exist in the project, and many of Wikipedia's internal disputes and conflicts in some way involve the delegation of work to an automated agent. Approved bots are explicitly delegated specific epistemic and organizational tasks by the Wikipedian community of volunteer editors, and bot developers are generally expected to be responsive to reasonable requests and concerns from the community.

However, it is important to note that nowhere in these two vignettes was it relevant for me to discuss how the source code of every single piece of software mentioned is publicly available for review. The ability to look into the source code of these bots, software tools, and even the underlying MediaWiki platform made little difference in the case of the professor interested in editing Wikipedia or in my case of trying to save the SaveMLAK article. What distinguishes myself as a veteran from a newcomer is not my access to the specific lines of code that specify how a particular browser extension automates a particular task, but rather how much expertise and familiarity I have working with these kinds of processes in this particular organizational culture. The professor would likely have been even more overwhelmed if I had also told him to read through AnomieBot's source code. I say this not to minimize the benefits of open sourcing socially important software code in a community, of which there are many. The relatively open development and approval process around both platform and bot/bespoke code in Wikipedia is a model that other platforms and institutions should closely examine. Yet it shows that "opening up the black box" is not a cure-all for the many complex issues that are being raised with the algorithmic systems being developed and deployed today in major media, communications, economic, and governmental platforms. In fact, one of the paradoxes of openness is that it can take substantial time, energy, investment, and resources to fully take advantage of all the materials released, as Tkacz (2015) notes in his analysis of participation in Wikipedia and other open platforms.

## Conclusion

The bots and other bespoke code that make up a core part of Wikipedia's culture are more than algorithmic systems with values embedded in them. They are a kind of social–technical design, ways in which the default affordances of wiki pages were intentionally extended by Wikipedians to support particular kinds of governance work in certain ways and not others. While these bots do increase the usability of the specialized page where such decision-making occurs, these efforts are about far more than simply making wikis in general more usable spaces for collaboration. In many cases, they make it far easier for veteran Wikipedians to engage in the kind of specific, complex, multifacted work involved in the governance of Wikipedia. This can make it far more difficult for newcomers to participate—not necessarily because bots, algorithms, or assisted tools are inherently difficult to deal with, but rather because bots support more complex kinds of governance practices in Wikipedia, and complex governance practices are difficult for newcomers. However, it is also important to understand why increasingly complex governance practices were developed Wikipedia starting around 2004, which is a much longer story.

Each of Wikipedia's clerk bots and assisted editing tools is developed with a different vision of what meta-level work in Wikipedia is and ought to be, as well as how this work is to be supported through automation. Some smaller and less controversial processes have remained relatively stable for years and involve only a single bot, like the edit request process supported by AnomieBot, or the Mediation Committee's process for dispute resolution, also supported by AnomieBot. The edit request process and the Mediation Committee are quite different than the deletion process, for example, which has been supported by 39 different bots that have assisted with various tasks at different times in the 10 years that this process has existed in Wikipedia. The infrastructures built to support these specialized processes undergird a wide range of different, simultaneously operating understandings about what Wikipedia—as an encyclopedia, community, organization, bureaucracy, public, institution, project, or any number of other mass nouns—is and ought to be. My analysis extends Tkacz's argument that studying the project's various specialized processes and venues "doesn't allow one to locate in Wikipedia a new organizational archetype; there is no generalizable Wikiocracy. Rather, it is the singularity of different organizational forms that such an approach accentuates" (Tkacz, 2015).

Those involved in newcomer socialization and mentorship have quite different assumptions, priorities, and goals compared to those involved in fast paced quality control, as can be seen in how these Wikipedians have differently developed and deployed bots in their spaces. In using bots and other bespoke code to extend the functionality of particular wiki pages inside of the broader wiki platform, these bot developers (and non-developers who participate in the development process in various ways) have made different decisions about not only what kinds of work ought to take place in Wikipedia, but how that work ought to be accomplished. Following Seaver's call (2017, this issue), in taking this kind of situated, contextual view, we can see such algorithms as culture, which opens up many new research topics beyond the important issues about the values embedded in algorithmic systems or the effects and impacts that such systems have.

Future research into algorithmic systems as culture should investigate Wikipedia's rich history of governance through automation and governance of automation. In a culture that highly values open source, there can be intense negotiation, reconfiguration, differential interpretation, and conflict across all layers of this "stack." In this context, scholarship that is based on identifying values embedded in the code of algorithmic systems only takes us so far, particularly when veterans in these organizational cultures have long made such critiques to each other internally. For example, how do the Wikipedians who believe that speedy deletion is too speedy contest the dominant algorithmic system? What about the Wikipedians who believe that newcomers should be able to participate without knowing templates and wiki code? And what about Wikipedians and Wikimedia Foundation staff who create new forms of automation to support alternative populations and priorities, like those supporting mentoring and socialization work (e.g., Halfaker et al., 2014)? Such topics and cases give scholars from across the disciplines a rich window into how algorithmic systems are actively developed and deployed alongside sociocultural and organizational structures.

## Acknowledgments

I would like to thank many people for their advice, feedback, and support on many iterations of this article and this research, including: Aaron Halfaker, Anna Lauren Hoffmann, Charlotte Cabasse-Mazel, Coye Cheshire, Deirdre Mulligan, Elizabeth Dubois, Gina Neff, Guilherme Sanches de Oliveira, Heather Ford, Jason Oakes, Jenna Burrell, Morgan Ames, Nick Seaver, Paul Duguid, Phil Howard, Rachel Bergman, Robyn Caplan, Sam Shorey, Sam Woolley, Tarleton Gillespie, the anonymous peer reviewers, and many Wikipedians and Wikimedia Foundation staff.

## Declaration of conflicting interests

I have previously been employed by the Wikimedia Foundation (WMF, the non-profit organization that operates Wikipedia) as a summer research intern in 2011 and as a research consultant from 2011–2012. This research project conceptually extends previous work I have done while employed by the WMF, but I have received no funding or other compensation from the WMF for the research project presented in this article. The WMF has made available a free and public computing infrastructure for researchers, including Toollabs (now Toolforge) and Quarry, which I used to generate various statistics.

## Funding

The author(s) disclosed receipt of the following financial support for the research, authorship, and/or publication of this article: This work was funded in part by the Gordon and Betty Moore Foundation (Grant GBMF3834) and by the Alfred P. Sloan Foundation (Grant 2013-10-27) to the University of California, Berkeley as part of the Moore-Sloan Data Science Environments. I am also grateful to the Computational Propaganda Project at the Oxford Internet Institute, the Algorithms in Culture project at UC-Berkeley, and the Berkeley Center for Law & Technology for funding various workshops and conferences where I received substantial feedback on this work. I also thank Yuvi Panda for our many conversations about these topics and for the Quarry project – a Wikimedia-hosted public service for querying Wikipedia's databases – which I used to generate several statistics presented in this article.

## Notes

1. Composite vignettes are common in anthropology and ethnography (Murchison, 2010). I have had this kind of interaction with several different people who have Wikipedia articles about themselves and/or their organizations, who came to me for help and found the edit request process difficult and confusing. Some details have been altered and merged from these multiple interactions to the presentation of this particular account.
2. A Google Scholar search in April 2017 returns 1350 publications that include a link to the official Wikimedia Foundation database dumps.
3. See data at: https://quarry.wmflabs.org/query/20703 for all edits (including discussion pages) and https://quarry.wmflabs.org/query/20704 for edits to articles.